\documentclass[12pt]{iopart}
\usepackage{graphicx}
\usepackage{iopams}
\usepackage{amssymb}
\usepackage{multirow}
\usepackage{xcolor}
\usepackage{soul}

\graphicspath{/}

\begin{document}
\title{$s_\pm+d$ wave multiband Eliashberg theory for the iron pnictides}
\author{G.A. Ummarino}
\ead{giovanni.ummarino@polito.it}
\address{Istituto di Ingegneria e Fisica dei Materiali,
Dipartimento di Scienza Applicata e Tecnologia, Politecnico di
Torino, Corso Duca degli Abruzzi 24, 10129 Torino, Italy; Department of Semiconductor Quantum Electronics / N.G.Basov High School of Physicists / Institute of Engineering Physics for Biomedicine, National Research Nuclear University MEPhI, Moscow Engineering Physics Institute, Kashira Hwy 31,
Moskva 115409, Russia}

\author{D. Torsello}
\ead{daniele.torsello@polito.it}
\address{Dipartimento di Scienza Applicata e Tecnologia, Politecnico di
Torino, Corso Duca degli Abruzzi 24, 10129 Torino, Italy; Istituto Nazionale di Fisica Nucleare, Sezione di Torino, Torino 10125, Italy}
\begin{abstract}
We calculated the critical temperature in the framework of $s_\pm+d$-wave multiband Eliashberg theory. We have solved these equations numerically to see at what values of the input parameters there is a solution with a non-zero critical temperature and what is the symmetry of the order parameter of this solution. For our model we consider the pnictide case with simplifications that allow us to obtain the most general possible information. For selected and representative cases in which the order parameter has $s_\pm+d$ symmetry, we calculated the superconducting density of states, the temperature dependence of the gaps, and the superfluid density so that comparison with experimental data can be made.
Finally, we show that such a system has only a twofold in-plane symmetry and undergoes a transition from nodal to fully gapped with increasing temperature.
\end{abstract}
%
%
\maketitle
\section{INTRODUCTION}
In the last thirty years, the study of superconducting materials had an astounding development. The starting point was ordinary low-temperature superconductors with a electron-phonon mechanism, a conducting band and isotropic order parameter ($s$ wave). Then, high-temperature cuprate superconductors \cite{HTS} with a non-phonon mechanism and a non-isotropic order parameter ($d$-wave)\cite{HTSd,HTCSumma1} were first discovered followed by multiband phononic materials such as fullerenes \cite{full,Fullumma2} and magnesium diboride \cite{MgB2,MgB2umma1,MgB2umma4}. Finally, iron-based compounds \cite{iron,Ironumma1,Ironumma2} appeared on the scene: multiband non-phononic materials with $s_\pm$ wave symmetry of the order parameter. The minimal model for the latter class of materials consists of only 2 conductivity bands (but even 5 bands were found to contribute to superconductivity, as in $LiFeAs$ \cite{LiFeAsumma}) with an isotropic order parameter with a phase difference of $\pi$ between each other \cite{sd}. In recent years, observations of possible mixed $s$ and $d-wave$ behaviour was proposed for non-phononic multiband superconductors \cite{divo1,divo2,divo3,divo4}, therefore we thought to develop, through Eliashberg theory, a possible general case considering in each band an order parameter with two components: one isotropic ($s$ wave) and one anisotropic ($d$ wave), realizing therefore a multiband $s+d$ wave where the two isotropic components ($s$ wave) are out of phase by $\pi$, called $s_\pm+d$. We consider the specific case of iron-based pnictide compounds, but our conclusions can be directly generalized to all systems in which the electron-boson coupling can be described in a similar way. 
In the following, we will write the Eliashberg equations for this new situation and we will see if there is a plausible range of physical input parameters (essentialy the electron-boson intra and inter band coupling constant $\lambda^{s,d}_{jk}$) where it is actually possible to have one or two order parameters with the two components ($s_\pm$ and $d$) at the same time. Then we will try to calculate physical observables that in the case ($s_\pm+d$) clearly differ from the pure $s_\pm$ and pure $d$ cases in order to propose experimental verifications for this model, and discuss situations in which they could occur.
\label{intro}
%
\section{MODEL: TWO-BAND ELIASHBERG EQUATIONS}
\label{sec:model}
We study a superconducting material with two conductivity parabolic bands (the simplest multiband case): we consider band one to be a hole band and band two an electron band. The cases with more than two bands can be reduced to effective two bands systems where the values of the coupling constant loose their physical meaning \cite{dolgov,torsello2019prb}, therefore this can be considered as a general scheme for multiband superconductors.
Our investigation starts from the consideration that, at the moment, the most studied multiband superconductors are the iron-pnictides where the mediation for the Cooper pairs is provided by antiferromagnetic spin fluctuactions, so this is the specific system we consider. Their lattice is described by the tetragonal symmetry at high temperature and by the orthorhombic symmetry in cryogenic conditions, which we study. In this case the isotropic part is repulsive (in the iron pnictides the contribution of phonons is very small \cite{Boeri} so we will put, in first approximation, this contribution equal to zero).
The electron-boson spectral functions (one for each band) have two components: one isotropic ($s$ wave) and one anisotropic ($d$ wave), yielding an overall anisotropic $s+d$ electron-boson interaction that is allowed only in the orthorhombic state. This happens because both the $s$- and $d$-wave states in the orthorhombic case belong to the same irreducible representation (A1g) \cite{Tsuei2000RMP}.
Solving the Eliashberg equations yields the superconducting gaps, and the more general solution has a two $s$ and two $d$ components (one for each band for each symmetry) where the two $s$ components are opposite in sign ($s_\pm$ wave), so the general solution for the order parameter is an $s_\pm+d$ wave. Both the $s$ and $d$ wave states are also compatible with a tetragonal symmetry of the system, whereas the mixed order can exist only in the orthorhombic phase. It is important to note that the \textbf{k}-dependent Eliashberg equations are nonlinear, for this reason the gaps can deviate from the symmetry of the interaction. Therefore, despite the symmetry of the interaction being fixed, that of the superconducting state is not. However the result must still be compatible with the orthorhombic state.
We calculated the experimental critical temperatures and the superconducting gaps by solving the $s_\pm+d$-wave two-band Eliashberg equations \cite{ummarinorev,Dwave1,Dwave2,Dwave5,Dwave6,ummarinoiron}. In this case, four coupled equations for the gaps $\Delta^{s,d}_{k}(i\omega_{n})$ and four for the renormalization functions $Z^{s,d}_{k}(i\omega_{n})$ have to be solved self consistently ($\omega_{n}$ denotes the Matsubara frequencies and $k=1,2$ the band index). The $s_\pm+d$-wave two-band Eliashberg equations (assuming that the Migdal theorem works \cite{Migdal}) in the imaginary axis representation, and in the compact shape (where the $s$ and $d$ components are not separated) read:

\begin{eqnarray}
\omega_{n}Z_{k}(\omega_{n},\phi)=\omega_{n}+\pi T
\sum_{j,m}\int_{0}^{2\pi}\frac{d\phi'}{2\pi}\Lambda_{kj}(\omega_{n},\omega_{m},\phi,\phi')N^{Z}_{j}(\omega_{m},\phi')
\end{eqnarray}
\begin{eqnarray}
&&Z_{k}(\omega_{n},\phi)\Delta_{k}(\omega_{n},\phi)=\pi T
\sum_{j,m}\int_{0}^{2\pi}\frac{d\phi'}{2\pi}[\Lambda_{kj}(\omega_{n},\omega_{m},\phi,\phi')-\mu^{*}_{kj}(\phi,\phi')
\big]\times\nonumber\\
&&\times\Theta(\omega_{c}-|\omega_{m}|)N^{\Delta}_{kj}(\omega_{m},\phi')
\label{eq:EE2}
\end{eqnarray}
where $\Theta(\omega_{c}-\omega_{m})$ is the Heaviside function, $\omega_{c}$ is a
cut-off energy and
\begin{eqnarray}
\Lambda_{kj}(\omega_{n},\omega_{m},\phi,\phi')=2\int_{0}^{+\infty}\Omega d\Omega
\alpha^{2}F_{kj}(\Omega,\phi,\phi')/[(\omega_{n}-\omega_{m})^{2}+\Omega^{2}]
\end{eqnarray}
\begin{eqnarray}
N^{Z}_{j}(\omega_{m},\phi)=
\frac{\omega_{m}}{\sqrt{\omega^{2}_{m}+\Delta_{j}(\omega_{m},\phi)^{2}}}
\end{eqnarray}
\begin{eqnarray}
N^{\Delta}_{j}(\omega_{m},\phi)=
\frac{\Delta_{j}(\omega_{m},\phi)}{\sqrt{\omega^{2}_{m}+\Delta_{j}(\omega_{m},\phi)^{2}}}
\end{eqnarray}
We assume \cite{ummarinorev,Dwave1,Dwave2,Dwave5,Dwave6} that the electron-boson spectral
functions $\alpha^{2}(\Omega)F_{kj}(\Omega,\phi,\phi')$ and the Coulomb
pseudopotential $\mu^{*}_{kj}(\phi,\phi')$ at the lowest order to contain
separated $s$ and $d$-wave contributions,
\begin{equation}
\alpha^{2}F_{jk}(\Omega,\phi,\phi')=\lambda^{s}_{jk}\alpha^{2}F^{s}_{jk}(\Omega)
+\lambda^{d}_{jk}\alpha^{2}F_{d}(\Omega)cos(2\phi)cos(2\phi')
\end{equation}
\begin{equation}
\mu^{*}_{jk}(\phi,\phi')=\mu^{*s}_{jk}
+\mu^{*d}_{jk}cos(2\phi)cos(2\phi')
\end{equation}
as well as the self energy functions:
\begin{equation}
Z_{k}(\omega_{n},\phi)=Z^{s}_{k}(\omega_{n})+Z^{d}_{k}(\omega_{n})cos(2\phi)
\end{equation}
\begin{equation}
\Delta_{k}(\omega_{n},\phi)=\Delta^{s}_{k}(\omega_{n})+\Delta^{d}_{k}(\omega_{n})cos(2\phi)
\end{equation}
The spectral functions $\alpha^{2}F^{s,d}_{jk}(\Omega)$ are normalized in the way that $2\int_{0}^{+\infty}\frac{\alpha^{2}F^{s,d}_{jk}(\Omega)}{\Omega}d\Omega=1$
In the more general case $\Delta_{k}(\omega,\phi')$ has $s$ and $d$ components while
the renormalization function $Z_{k}(\omega,\phi')=Z^{s}_{k}(\omega)$ has just the $s$ component because
the equation for $Z^{d}_{k}(\omega)$ is a homogeneous integral
equation whose only solution in the weak-coupling regime is
$Z^{d}_{k}(\omega)=0$ \cite{zetad}. For simplicity we also assume that
$\alpha^{2}F_{s}(\Omega)=\alpha^{2}F_{d}(\Omega)$ and that the
spectral functions is the difference of two Lorentzian, i.e.
$\alpha^{2}F^{s,d}_{kj}(\Omega)=C^{s,d}_{kj}[L(\Omega+\Omega^{s,d}_{0,kj},\Upsilon^{s,d}_{kj})-L(\Omega-\Omega^{s,d}_{0kj},\Upsilon^{s,d}_{kj})]$
where $L(\Omega\pm\Omega^{s,d}_{0,kj},\Upsilon^{s,d}_{kj})=[(\Omega\pm\Omega^{s,d}_{0,kj})^{2}+(\Upsilon^{s,d}_{kj})^{2}]^{-1}$,
$C^{s,d}_{kj}$ are the normalization constant necessary to obtain the proper
values of $\lambda^{s,d}_{jk}$, $\Omega^{s,d}_{0jk}$ and $\Upsilon^{s,d}_{jk}$ are the peak
energies and half-width, respectively. This choice of the shape of spectral function is a good approximation of the true spectral function connected with antiferromagnetic spin fluctuations \cite{ummarinoiron}.
In all the calculations we set $\Omega^{s,d}_{0,kj}=\Omega_{0}$, i.e. we assume that the characteristic energy of spin fluctuations is a single quantity for all the coupling channels, and  $\Upsilon^{s,d}_{kj}= \Omega_{0}/2$, based on the results of inelastic neutron scattering measurements \cite{Inosov}.
The peak energy of the Eliashberg functions, $\Omega_0$, can be directly associated to the experimental critical temperature, $T_c$, by using the empirical law $\Omega_{0}=2T_{c}/5$ that has been demonstrated to hold, at least approximately, for all iron pnictides \cite{Paglione}. We have chosen a critical temperature $T_{c}=32$ K which is a typical value for iron pnictides \cite{divo1,divo2}. It is important to note that the choice of these spectral functions is what makes this model specific for the iron pnictides, and that these results are valid for any system for which the electron-boson interaction can be described in this way.
In the first approximation we put $\mu^{*}_{jk}(\phi,\phi')=0$ \cite{Mazincoulomb} and we do not include disorder induced scattering that is negligible for high quality materials  without artifically introduced defects \cite{torsello2020prappl,torsello2021scirep,torsello2020sust,ghigobook}.

We solve the imaginary axis $s_\pm+d$-wave two-band Eliashberg
equations for different values of $\lambda^{s,d}_{ij}$ and calculate the critical temperature and the symmetry of superconducting gaps. Then, via Pad\`{e} approximants \cite{Vidberg}, we calculate the low-temperature value
($T<T_{c}/10$) of the gaps because, in presence of a strong coupling
interaction, the value of $\Delta^{s,d}_{k}(i\omega_{n=0})$ obtained by solving the imaginary-axis
Eliashberg equations can be very different from the value of
$\Delta^{s,d}_{k}$ obtained from the real-axis Eliashberg equations.
After determining the gaps as a function of energy, we will be able to calculate the densities of the superconducting states ($DOS$) which is directly comparable with tunneling measurements at very low temperatures.
We calculate also the temperature dependence of the order parameters in the $s_\pm+d$ case and the corresponding superfluid density.
We use the following model for the values of the coupling constant (the input parameters) $\lambda^{k}_{ij}=
\left(
  \begin{array}{cc}
    \lambda^{s,d}_{11}=0.1+n|\lambda^{s,d}_{12}| & \lambda^{s,d}_{12} \\
    \lambda^{s,d}_{21}=\lambda^{s,d}_{12} & \lambda^{s,d}_{22}=0.1+0.5n|\lambda^{s,d}_{12}| \\
  \end{array}
\right)$
with $i,j=1,2$,$n\geq 0$, $\lambda^{s}_{12} < 0$ and $\lambda^{d}_{12} > 0$.
Of course, $\lambda^{s}_{12} > 0$ is not possible without phonons (standard $s+d$ case) so we neglect it.
We solve the Eliashberg equations in three different cases: $n=0$ (pure interband case), $n=1$ (all $\lambda_{ij}$ are comparable in size among them) and $n=10$ (weak interband coupling $\lambda_{ij}<< \lambda_{ii}$). At the moment we study just the situation
where the value of $n$ is the same for the two groups of $\lambda^{s}_{ij}$ and $\lambda^{d}_{ij}$.
\section{RESULTS AND DISCUSSION}

\begin{figure}[ht]
\begin{center}
\includegraphics[keepaspectratio, width=8cm]{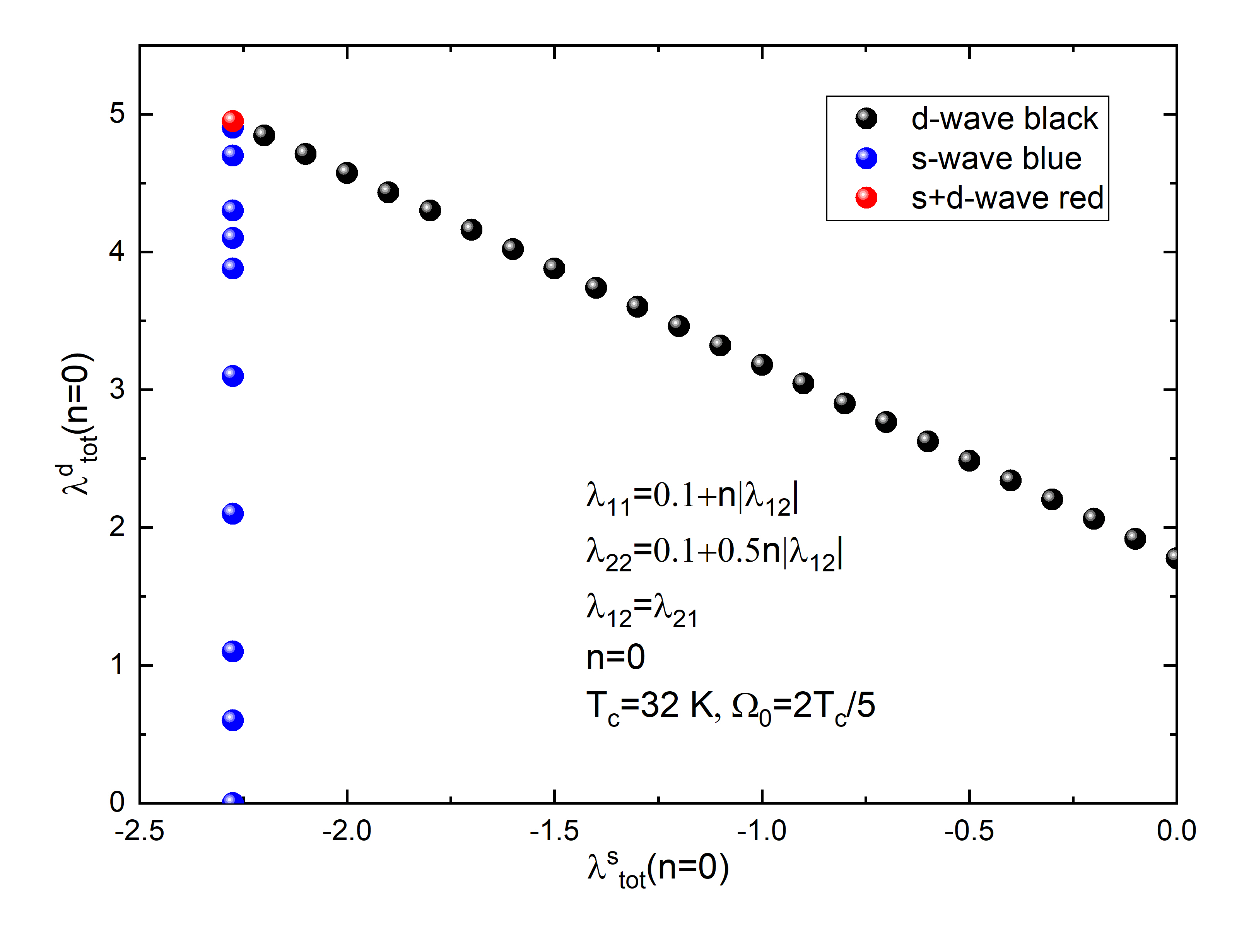}
\caption{Phase diagram of the order parameter symmetry with varying total coupling constants $\lambda^{d}_{tot}$ and $\lambda^{s}_{tot}$, $n=0$ case (pure interband).}\label{fig:phase1}
\end{center}
\end{figure}

\begin{figure}[h]
\begin{center}
\includegraphics[keepaspectratio, width=8cm]{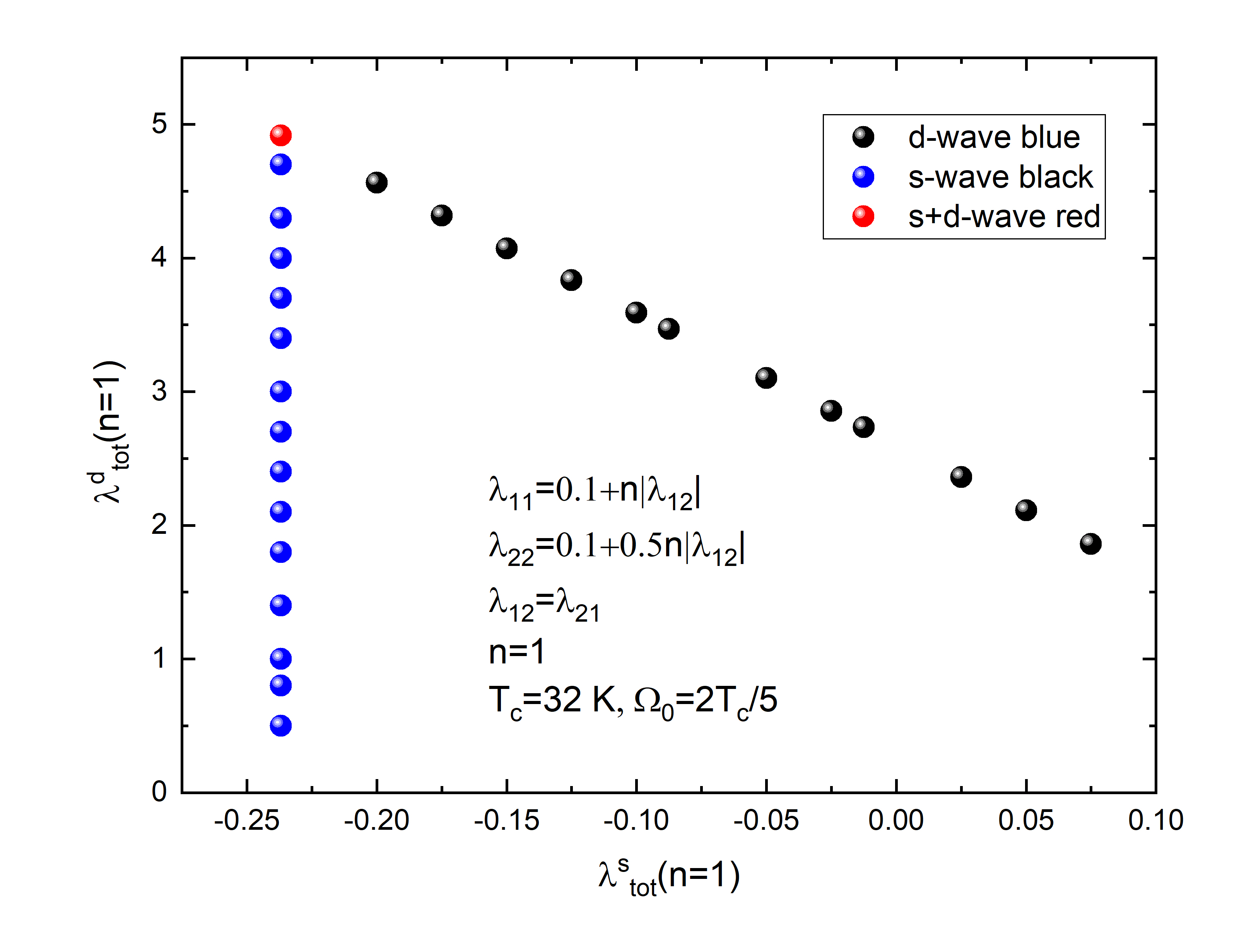}
\caption{Phase diagram of the order parameter symmetry with varying total coupling constants $\lambda^{d}_{tot}$ and $\lambda^{s}_{tot}$, $n=1$ case (balanced).}\label{fig:phase2}
\end{center}
\end{figure}

\begin{figure}[h]
\begin{center}
\includegraphics[keepaspectratio, width=8cm]{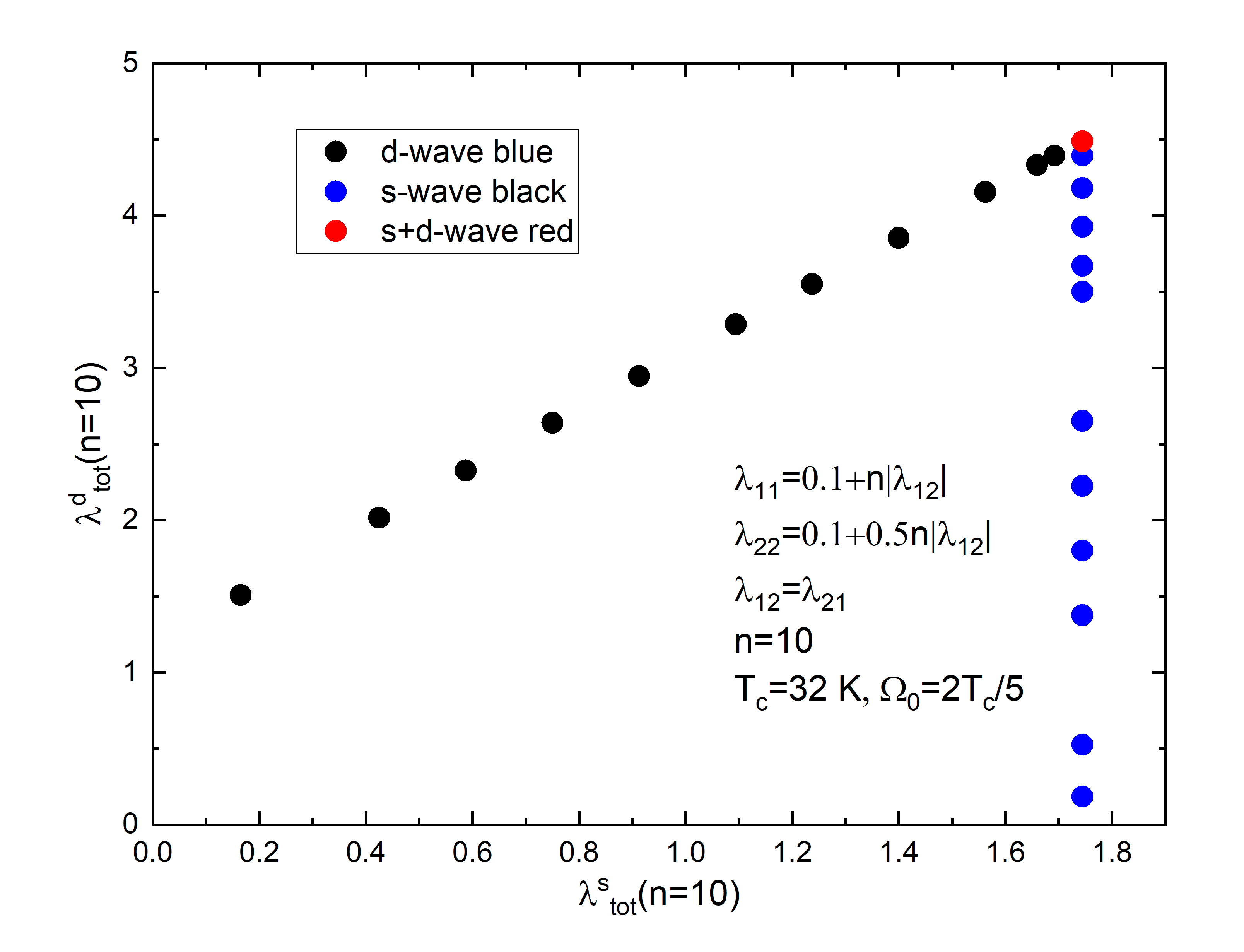}
\caption{Phase diagram of the order parameter symmetry with varying total coupling constants $\lambda^{d}_{tot}$ and $\lambda^{s}_{tot}$, $n=10$ case (pure intraband).}\label{fig:phase3}
\end{center}
\end{figure}

Figures \ref{fig:phase1}, \ref{fig:phase2} and \ref{fig:phase3} show the phase diagrams of this model with the symmetries of the order parameters as functions of the two total coupling parameters ($\lambda^{s}_{tot}$ and $\lambda^{d}_{tot}$) (where $\lambda^{s,d}_{tot}=\sum_{i,j=1}^2 N_i(0)\lambda^{s,d}_{ij}/\sum_{i=1}^2 N_i(0)$, and $N_i(0)$ is the normal density of states at the Fermi level for the $i$ band) for the three $n$ values discussed above.
We see that in all cases there are only a couple of values of $\lambda^{s,d}_{totmax}$ at which the $s_\pm+d$ symmetry is realized. This occurs at the intersection of the stability regions of $s-$ and $d-$ superconductivity. It should be noted that, as visible from the phase diagrams, only a narrow subset of parameters gives a self-consistent solution of the Eliashberg equations. Outside of these regions the calculations yield a critical temperature different from the one that was set (and that determines $\Omega_0$): lower for smaller coupling parameters and higher for larger ones. Therefore, such results need to be discarded because are not consistent with the model assumptions. Full calculations for systems with different $T_c$ result in equal curves. The reason for this lies in the fact that we have imposed a further constraint (deduced from the experimental data) between $\Omega_0$ and $T_c$.\\

\begin{figure}[h!]
\begin{center}
\includegraphics[keepaspectratio, width=8cm]{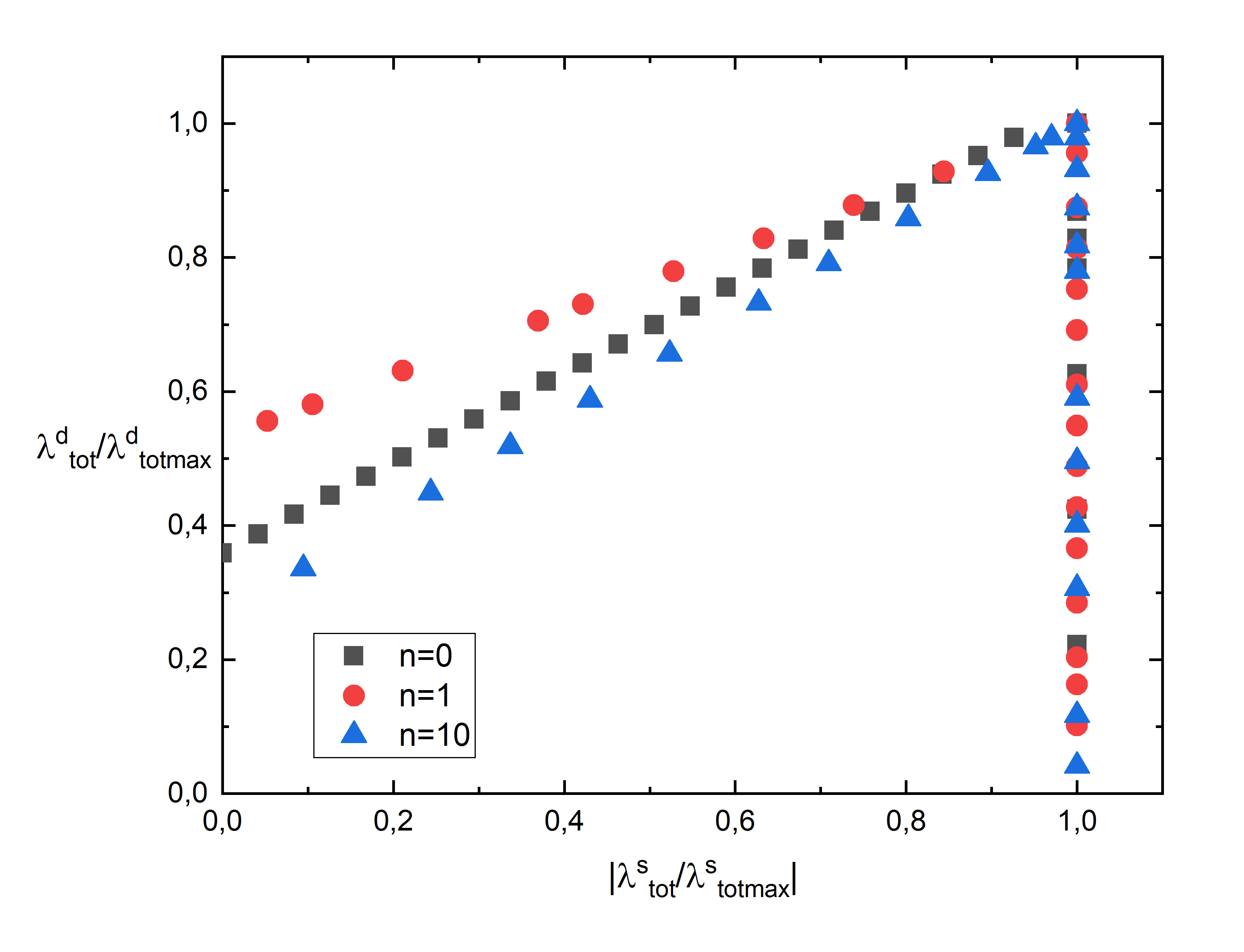}
\caption{Phase diagram of the order parameter symmetry with varying normalized total coupling constants: $n=0$ black squares, $n=1$ red circles and $n=10$ dark blue up triangles}\label{fig:phase4}
\end{center}
\end{figure}

The three cases normalized to the maximum values of $\lambda^{s,d}_{tot}$ are shown together in figure \ref{fig:phase4}. The behaviour is similar, but not quite the same. This shows that such $s_\pm+d$ symmetry of order parameter can be realized in different systems (with predominant interband, intraband or balanced character), but in a very narrow parameter space in which the two symmetry contributions balance each other. 
It is important to remind here that the admixture of the $s$- and $d$- states is allowed in the orthorhombic state, as well as the pure cases, because both states belong to the same irreducible representation ($A_{1g}$). However, when one of the contributions ($s$ or $d$) is suppressed a sort of "symmetry elevation" with respect to the interaction seems to take place, as observed in other fields \cite{Karlsson2010PRB,Dupertuis2011PRL}.\\


\begin{figure}[h!]
\begin{center}
\includegraphics[keepaspectratio, width=8cm]{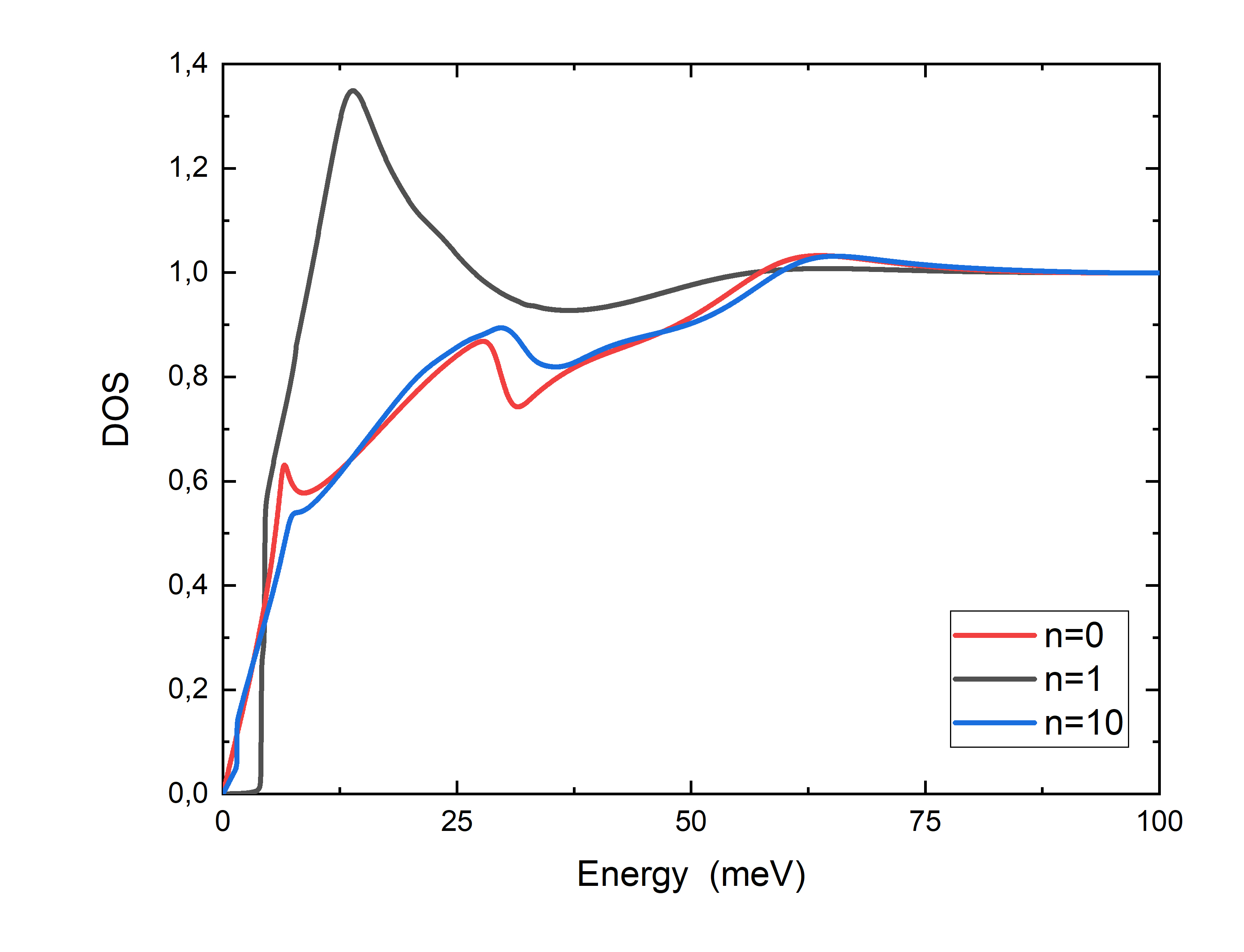}
\caption{Densities of states (DOS) at very low temperatures ($T=1.5 $K) for the order parameter of the $s_\pm+d$ symmetry as functions of energy in the three different cases: $n=0$ red solid line, $n=1$ black solid line and $n=10$ dark blue solid line.
}\label{fig:dos}
\end{center}
\end{figure}

The low temperature densities of states as a function of energy for the $s_\pm+d$ symmetry are shown in figure \ref{fig:dos}.
 We show three different situations: $n=0$, $n=1$ and $n=10$ (correspondent to intraband, balanced and interband cases, respectively). The energy dependences of the DOS in the cases $n=0$ and $n=10$ are similar and unusual, whereas the case $n=1$ is different and more standard. It is clearly seen that the form of the energy dependence of the $DOS$ in the case when the order parameter has $s_\pm+d$ symmetry ($n=0$ and $n=10$) has rather unusual characteristics that differ profoundly from the $s$- and $d$-cases, and can be easily identified in tunneling experiments.\\

\begin{figure}[h!]
\begin{center}
\includegraphics[keepaspectratio, width=\linewidth]{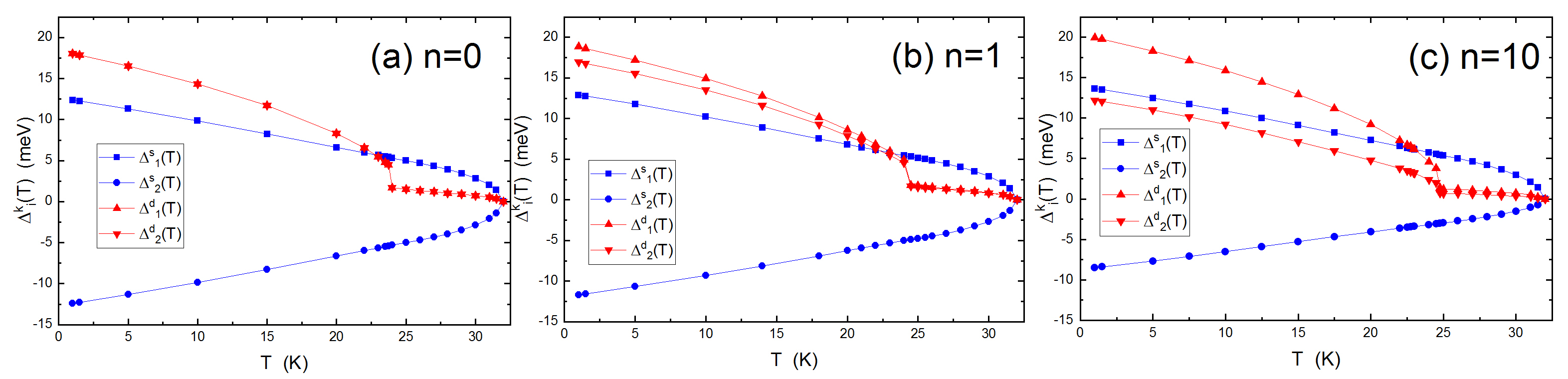}
\caption{Temperature dependences of the order parameters of the $s_\pm+d$ symmetry calculated by solution of Eliashberg equations on imaginary axis for the three different cases: $n=0$ panel $(a)$, $n=1$ panel $(b)$ and $n=10$ panel $(c)$ The dark blue symbols are the $s$-wave component, the red symbols are the $d$-wave components.}\label{fig:gap}
\end{center}
\end{figure}

 The temperature dependences of the order parameter of the $s_\pm+d$ symmetry, calculated by solving the Eliashberg equations on the imaginary axis for three different cases, are shown in figure \ref{fig:gap} (panels $a, b, c$). It can be seen that there is a jump in the behavior of the d-component of the order parameter around $T^*\sim 24$ K. This happens when the largest $d$-wave gap becomes smaller than the smallest $s$-wave one. Moreover, the values of $2\Delta^{s,d}_{i}/k_{B}T_{c}$ are much larger than the $BCS$ value.\\

\begin{figure}[h!]
\begin{center}
\includegraphics[keepaspectratio, width=8cm]{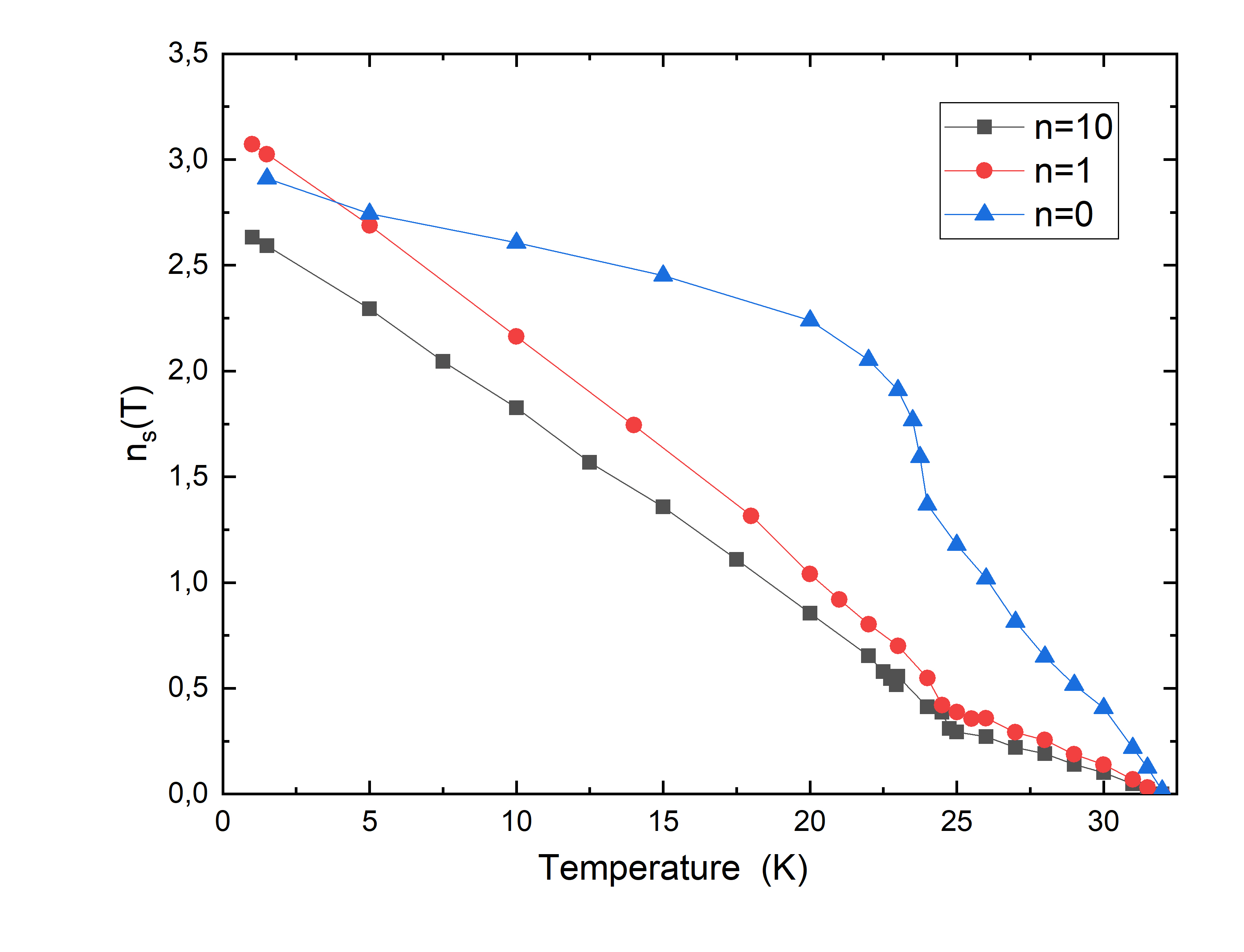}
\caption{Temperature dependence of the superfluid density in the system with $s_\pm+d$ symmetry of the order parameter, calculated by solution of Eliashberg equations on imaginary axis for the three different cases: $n=0$ dark blue up triangles, $n=1$ red circles and $n=10$ black squares.}\label{fig:rho}
\end{center}
\end{figure}

\begin{figure}[h!]
\begin{center}
\includegraphics[keepaspectratio, width=8cm]{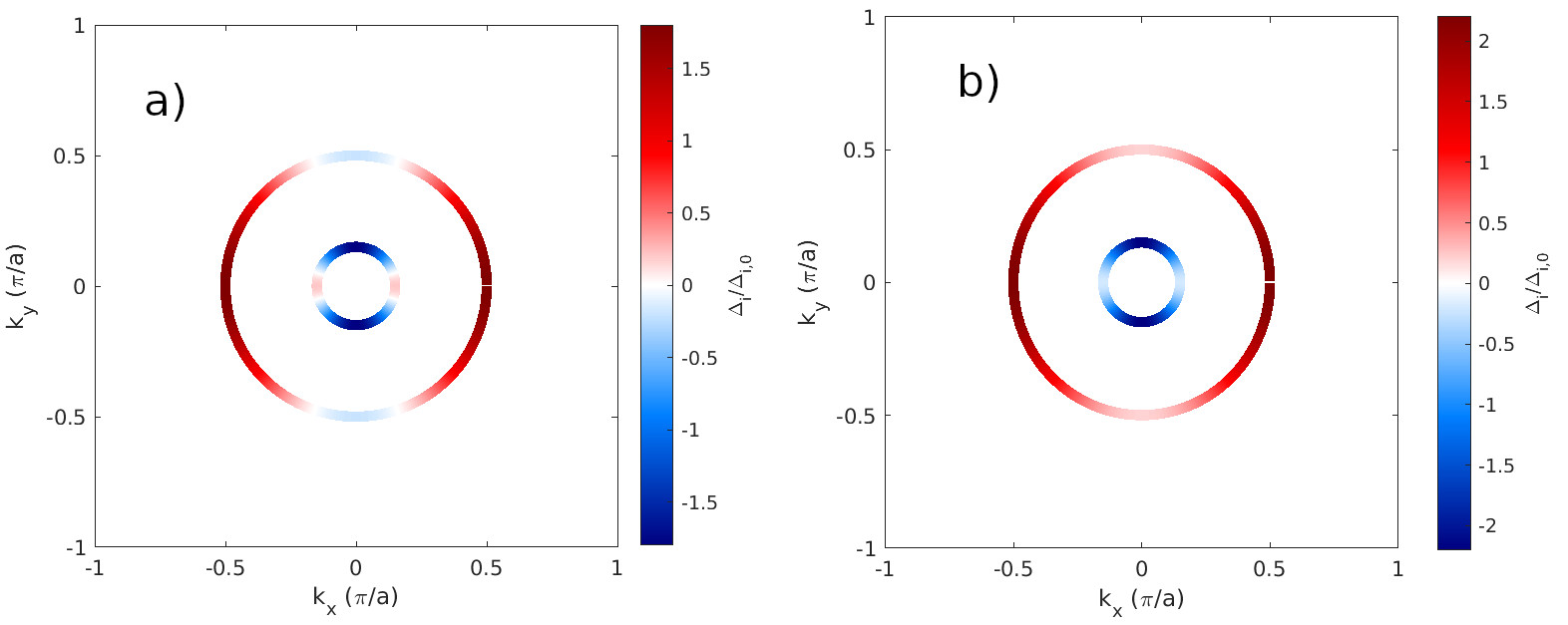}
\caption{Schematic view of the gap values over a typical two band tubular Fermi surface (FS) folded in the Brillouin zone. Panel (a) shows the FS of the system where $\Delta_d > \Delta_s$ while panel (b) shows the FS of the system where $\Delta_s > \Delta_d$.}\label{fig:FS}
\end{center}
\end{figure}

 The temperature dependence of the superfluid density in a system with $s_\pm+d$ symmetry of the order parameter, calculated by solving the Eliashberg equations on the imaginary axis for three different cases, is shown in Figure \ref{fig:rho}. There is a jump in the temperature dependences of the superfluid density at the same temperatures as for gaps of pure $d$-wave symmetry.
 The cases $n=1$ and $n=10$ are similar, while the case $n=0$ is completely different.\\
 At last, we discuss what such a gap structure would look like on the Fermi surface ($FS$) of a real material. In Figure \ref{fig:FS}, the magnitude of the gap (shown by the colormap) is represented on a generic Fermi surface consisting of two tubular Fermi sheets in the first Brillouin zone for two cases: when the $d$-wave gap value exceeds the $s$-wave gap (panel $a$, below $T^*$) and for the opposite situation (panel $b$, above $T^*$). It is clear the tetragonal symmetry is lost that in both cases and only twofold symmetry is retained, as occurs in the nematic state \cite{Carretta}. Moreover, the order parameter is nodal only at low temperature, and the system shows a transition to a fully gapped order parameter at $T^*$. This pecularity should be visible in tunneling measurements and with spectroscopic techniques.

\label{sec:results}
\section{CONCLUSIONS}
We have solved the two band Eliashberg equations spanning the coupling constants parameter space in order to understand which pairing symmetries are possible if both $s_\pm$ and $d$ wave coupling are considered.  We considered the specific case of iron-based pnictide compounds, but our conclusions can be directly generalized to all orthorhombic systems in which the electron-boson coupling can be described in a similar way.  For a fixed critical temperature value and balanced bands, only a pair of $\lambda^{s}_{tot}$ and $\lambda^{d}_{tot}$ input parameters yields the $s_\pm+d$ mixed symmetry. All other stable solutions lose one of the two contributions ($s$ or $d$), showing a sort of symmetry elevation with respect to the electron-boson interaction, a fact allowed by the nonlinearity of the problem and by the compatibility with the orthorhombic state. For the $s_\pm+d$ cases, we calculated observables such as the density of superconducting states at low temperature and the temperature dependence of the superfluid density in order to make theoretical predictions that can be compared with experiments. In this way it will be possible to determine if this symmetry is really present in some of the systems where it has been proposed. Finally, we discussed the properties of the order parameter in terms of symmetry and nodality, highlighting a transition from a nodal to a fully gapped order parameter at a finite temperature $T^*$.
\label{sec:conclusions}
\ackn
This work was supported by the MEPhI Academic Excellence Project (contract No. 02.a03.21.0005) for the contribution of prof. G.A. Ummarino.
D. Torsello acknowledges the partial support by the Italian Ministry of Education, University, and Research (Project PRIN HIBiSCUS, Grant No. 201785KWLE) and “Programma Operativo Nazionale (PON) Ricerca e Innovazione 2014–2020”. The authors are extremely thankful to D. Daghero for fruitful discussion.\\

%
%
%
%
%
%
\end{document}